\documentclass[12pt]{article}

\usepackage{amsfonts}
\usepackage{amsthm}
\usepackage{amsmath}
\usepackage{graphicx}
\usepackage{bourgain}
\usepackage{hyperref}
\newcommand{\remove}[1]{}

\newcommand\set[1]{{\{{#1}\}}}

\newcommand{\perm}{\mathsf{permanent}}

\newcommand{\gnote}[1]{}
\newcommand{\nnote}[1]{}
\renewcommand{\Ex}{\mathbb{E}}
\newcommand{\inner}[1]{{\langle{#1}\rangle }}

\begin{document}

\title{Gaussian Noise Sensitivity and BosonSampling\thanks{This work was carried
 out in part while the authors were visiting the Simons Institute for the Theory
 of Computing at UC Berkeley.}}

\author{Gil Kalai\thanks{Einstein Institute of Mathematics, the Hebrew University
  of Jerusalem, and Department of Mathematics, Yale University. Supported by an
ERC grant and by an NSF grant.} \and
Guy Kindler\thanks{School of Computer Science and Engineering,
    the Hebrew University. Supported by an Israeli  Science   Fund
    grant and a Binational Science Fund grant no.\ 2008477.}
  }

\maketitle

\begin{abstract}
We study the sensitivity to noise of $|\perm (X)|^2$ 
for random real and complex
$n \times n$ Gaussian matrices $X$, and show that asymptotically the correlation between the noisy and noiseless
outcomes tends to zero
when the noise level is $\omega(1)/n$. This suggests that, under certain reasonable noise models,
the probability distributions produced by noisy BosonSampling are very sensitive to noise. 
We also show that when the amount of noise is constant 
the noisy value of $|\perm (X)|^2$ can be approximated efficiently on a classical computer.
These results seem to  weaken the possibility of demonstrating quantum-speedup via BosonSampling without quantum fault-tolerance.

\end{abstract}

\thispagestyle{empty}
\newpage

\section {Introduction}

\paragraph {BosonSampling.}
BosonSampling (Aaronson and Arkhipov \cite {AaAr13}, see also Tishby and Troyansky \cite {TrTi96}) 
is the following computational task.
\vspace{0.1 in}

\begin {enumerate}
\item

The input is  an $n$ by $m $ complex matrix whose rows are unit vectors.

\item
The output is a sample from a probability distribution on all
multisets of size $n$ from $\{1,2,\dots ,m\}$, where the probability of a multiset $S$ is
proportional to $\mu(S)$ times the square of the absolute
value of the permanent of the associated $n$ by $n$ minor. Here, if the elements of the 
multiset occurs with multiplicities $r_1,r_2,\dots,r_k$, then $\mu(S)=1/r_1!r_2!\dots r_k!$.

\end {enumerate}

\vspace{0.1 in}

This sampling task can be achieved by an (ideal) quantum computer.
In fact, it can be realized by linear systems of $n$ noninteracting photons which
describe a restricted regime of quantum algorithms.
The analogous algorithmic task with determinants instead of permanents is
referred to as FermionSampling. While FermionSampling is in {\bf P}, a polynomial algorithm for
BosonSampling implies that the polynomial hierarchy collapses to the third level \cite{AaAr13}.

When we consider noisy quantum computers with the full apparatus of quantum fault-tolerance,
BosonSampling can be achieved with negligible error.  A few years ago, Aaronson and Arkhipov proposed
a way based on BosonSampling to demonstrate quantum speed-up without
quantum fault-tolerance\footnote{``quantum speed-up,'' ``quantum
  supremacy'' and ``falsification of the extended Church Turing
  Theses,'' are all terms used to express the hypothesis of
  computationally superior quantum computing.}  They conjectured that,
on the computational complexity side, achieving an approximate version
of BosonSampling, even for a (complex) Gaussian random matrix, will be
computationally hard for classical computers. On the other hand they
conjectured that such approximate versions can be achieved when the
number of bosons is not very large, but still large enough to
demonstrate ``quantum supremacy.''  


\paragraph{Noise sensitivity of Gaussian matrices.} An $n\times n$
complex (real) Gaussian matrix is a
matrix where the coordinates are independent and are chosen according to a
normalized Gaussian distribution. If $X$ is an $n\times n$ matrix and
$U$ is a Gaussian matrix, then the random matrix
$Y=\sqrt{1-\epsilon}\cdot X+\sqrt \epsilon U$ is called an
$\epsilon$-noise of $X$.

\begin{theorem}
\label{thm:1}
Let $X$ be an $n\times n$ random Gaussian complex (real) matrix,  let
$\epsilon>\omega\parenth{\frac1n}$, and let $Y$ be an $\epsilon$-noise
of $X$. Define
\[
f(X)= |\perm (X) |^2,~~ g(X)=\Ex \Brac{ |\perm(Y)|^2\ \vert X}.
\]
Then

(i) As long as $\epsilon = \omega (\frac{1}{n})$, the correlation between $f$ and $g$ tends to zero. 
In other words:

\begin {align}
\label{e:i}
corr(f,g)=\frac {<f',g'>}{\|f'\|_2\|g'\|_2} = o(1),
\end {align}

where $f'=f-\mathbb E(f)$ and 
$g'=g-\mathbb E(g)$. 



(ii) For $d \gg 1/\epsilon $ there is a degree $d$ polynomial function of $X$, $p_d(X)$,
such that
\begin {align}
\label {e:ii}
\norm{p_d(X)-g(X)}_2^2 = o( \norm{g}_2^2 ).
\end {align}

(iii) 
Moreover, any coefficients of $p_d$ can be computed
in polynomial time in $n$, and $p_d$ can also be approximated to
within a constant by a constant-depth circuit.
\end{theorem}

The proof of Theorem \ref {thm:1} for the real case relies on the
description of noise in terms of the Fourier-Hermite expansion.
The study of noise-sensitivity requires an understanding of how the $\ell_2$ norm is distributed
among the degrees in the Hermite expansion. As it turns out the contributions coming
from degree $2k$ coefficient is $ (k+1) (n!)^2$. The combinatorics involved is related to Aaronson and Arkhipov's 
computation of the forth moment of $|\perm(A)|$ when $A$ is a complex
Gaussian matrix. In the complex case, which is similar but somewhat
simpler, we use another set of orthogonal functions which form
eigenvectors of the noise operator. In this basis the contribution of
the degree $2k$ coefficients is $(n!)^2$ for all $k=0,1,\ldots, n$.

\medskip \noindent 
We also obtain fairly concrete estimates:

\begin {corollary}[of the proof]
\label{c:1}
For the complex case, 
\begin {align}
\label {e:c1-1}
corr(f,g)=\sqrt{  \frac{ (1-(1-\epsilon)^n)\cdot(2-\epsilon) }
  {\epsilon n\cdot (1+(1-\epsilon)^n)
  }  }
\end {align}

For $\epsilon = c/n$ this asymptotically gives  

\begin {align}  
\label {e:c1-2}
corr (f,g)
= \sqrt {\frac {2\cdot(1-e^{-c})}{c\cdot(1+e^{-c})} }.
\end {align}

\end {corollary}

\begin{figure}
\centering
 \includegraphics[scale=0.9]{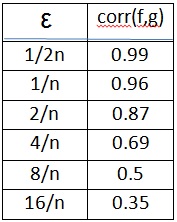}
 \caption{The correlation between the noisy and ideal values of the BosonSampling coefficients (for terms without repeated columns,) 
 for several values of noise.} 
 \label{fig:1}
 \end{figure}

See Figure \ref{fig:1} for some values. We also note that the
asymptotic values given there via formula (\ref{e:c1-2}) are quite
close to the values for small number of bosons $n=10, 20, 30$ as given
by \eqref{e:c1-1}.



\paragraph {Noise sensitivity of BosonSampling.}

Given an $n$ by $m$ matrix drawn at random from a (real or complex) Gaussian distribution,
we can compare the distribution of BosonSampling and of ``noisy BosonSampling'', where the later is described
by averaging over an additional $\epsilon$-noise. 





Theorem \ref{thm:1} suggests that for any fixed amount of noise
$\epsilon >0$, noisy BosonSampling can be approximated in ${\bf P}$ 
and that, as long that $\epsilon=\omega(\frac1n)$, the
correlation between BosonSampling and noisy BosonSampling tends to
$0$. We say ``suggests'' rather than ``asserts'', because when we move
from individual permanents to permanental distributions we face two
issues. The first is that averaging the probability of a minor is not
identical to averaging the value of permanent-squared: the latter does
not take into account the normalization term, which is the weighted sum of
squares of permanents for all $n$ by $n$ minors.\footnote{When the rows
of the matrix are orthonormal then the weighted sum of all permanents is 1. 
In the more general case we consider it is given by the Cauchy-Binet
theorem for permanents \cite {Min78,HCB88}.}  However, we can
expect that approximating the normalization term itself is in ${\bf P}$ for a fixed amount of noise,  
and that when $m$ is not too small w.r.t. $n$ the normalization
term will be highly concentrated so it will have a small effect.  The
second issue is that when $m$ is not too large w.r.t. $n$ a typical
permanent for BosonSampling will have repeated columns and this will
require an (interesting) extensions of our results, which is yet to be
done.  When $m$ is large compared to $n^2$ we will have that the
BosonSampling distribution is mainly supported on permanents without
repeated columns.

We also note that Theorem \ref {thm:1} and its consequences refer to
correlation 
between distributions rather than to the variational $(\ell_1$)
distance that Aaronson and Arkhipov discuss.  We expect that when the
amount of noise is $C/n$ then $f(x)$ and $g(x)$ are bounded away in
the $\ell_1$-distance by a constant depending on $C$ (This is
suggested but not implied by the 
correlation estimate of part (i) of Theorem \ref{thm:1}). We also
expect that for every $n$ and $m$ ($m \ge n$, say), when the amount of
noise is $C/n$ then the noisy BosonSampling distribution is bounded
away from the noiseless BosonSampling distribution in the $\ell_1$-distance.




While not proven here, we also expect that our results can be extended in the following three directions
\begin {enumerate}

\item
The results apply to other forms of noise like a deletion of $k$ of our $n$ bosons at random,
or modeling the noise based on the ``gates,'' namely the physical operations needed for
the implementation, or noise representing "incomplete interference."


\item The results about noisy permanents extend also to the case of repeated columns.


\item
Noise sensitivity extends to describe the sensitivity of the distribution under small perturbations of the noise parameters.  
%

\end {enumerate}

All in all Theorem \ref {thm:1} raises the question of whether, without quantum-fault-tolerance, approximate BosonSampling
in Aaronson and Arkhipov's sense is realistic  
and whether realistically modeled noisy BosonSampling 
manifests computational-complexity hardness. Noise sensitivity for squares of 
permanents and BosonSampling may be manifested even for realistic levels of noise  
even for small values of $n$ and $m$ 
(Say, 10 bosons with 20 modes.) To this end computer simulations can give a good picture, and, of course, experimental efforts for 
implementing BosonSampling for three, four, five, and six bosons may
also give us good picture on how things scale. This is discussed
further in Appendix 2.  

Studying noise sensitivity of other
quantum ``subroutines'' such as FourierSampling,  processes for creating anyons of various types, and tensor networks,
is an interesting subject for further study. 

We note also that there are various results in the literature both in the study of controlled quantum systems \cite {KKK14}   
and in computational complexity \cite {BL12,MMV13}, 
demonstrating that ``robustness'' and ``noise stability'' lead to computational 
feasibility\footnote {As the PCP theorem demonstrates this is not {\em always} the case.}

The structure of the paper is as follows: Section \ref {s:back} gives further background on BosonSampling and noise sensitivity.
The proof of  theorem \ref {thm:1} for complex Gaussian matrices is given in Section \ref {s:complex}, and for 
the real case is delayed to the appendix in Section \ref{s:real}. 
Section \ref {s:conc} has some further discussion interpreting our results,  and the appendices elaborate on several
extensions and related issues.


\section {Background}
\label{s:back}

\subsection {Noise sensitivity}

The study for noise sensitivity for Boolean functions
was introduced by Benjamini, Kalai, and Schramm \cite {BKS99}, see also \cite {GaSt14}.
The setting for Boolean functions on $R^n$ equipped with the Gaussian probability distribution
was studied by Kindler and O'Donnell \cite {KiOd12}, see also Ledoux \cite {Led96}, and O'Donnell \cite {O'Do14}.

Let $h_j(x)$ be the normalized Hermite polynomial of degree $j$.
For $d=(d_1,\ldots,d_n)$ we can define a multivariate Hermite
polynomial $h_d(X)=\prod_{i=1}^n h_{d_i}(x_i)$, and the set of such
polynomials is an orthonormal basis for $L_2(\R^n)$.

Let $f$ be a function from $\R^n$ to $\R$.
Let $\epsilon>0$ be a noise parameter and let
$\rho=\sqrt{1-\epsilon}$. We define $T_\rho(f)(x)$ to be the expected value of $f(y)$ where $y=\sqrt {1-\epsilon}x +\sqrt \epsilon u$,
and $u$ is a Gaussian random variable in $\R^n$ of variance 1.
Consider the expansion of $f$ in terms of Hermite polynomials

\begin{align}
f(x) = \sum _{\beta \in \N^d}\hat f(\beta ) \prod_{i=1}^d h_{\beta_i}(x_i).
\end{align}

 The values $\hat f(\beta ) $ are called the Hermite coefficients of $f$.
Let $| \beta |=\beta_1+\cdots +\beta_n$

The following description of the noise operator in terms of Hermite expansion is well known:

\begin{align}
\label{e:nf}
T_\rho (f) = \sum _{\beta \in N^d}\hat f(\beta )\rho^{\beta} \prod_{i=1}^d h_{\beta_i}(x_i).
\end {align}

A class of functions with mean zero $\cal F$
is called (uniformly) noise-stable if there is a function $s(\rho)$ that tends to zero with $\epsilon$
such that
for every function $f$ in the class,
\[
 \| T_\rho(f) - f\|_2^2 \le s(\rho)\|f\|_2^2.
\]
A sequence of function $(f_n)$ (with mean zero) is asymptotically noise-sensitive if for every $\epsilon>0$
\[
 \|T_\rho(f)|_2^2 = o(1)|f\|_2^2.
\]

These notions are mainly applied for characteristic functions of events (after subtracting their mean value).
There are several issues arising when we move to general functions.
In particular, we can consider these notions w.r.t. other norms.
Noise-stability is equivalent to the assertion that most of the $\ell_2$-norm of every $f \in {\cal F}$
is given by low-degree Hermite-coefficients. Noise sensitivity is equivalent to the assertion that the contribution of Hermite coefficients of low degrees
is $o(\|f\|_2^2)$.

{\bf Example:} Let $f$ be a function of $n^2$ (real) Gaussian variables describing the entries of an $n$ by $n$ matrix,
given by the permanent of the matrix.
In this case the $n!$-terms expansion of the permanent is its Hermite expansion.
This gives that the expected value of the permanent squared is $n!$.
The permanent is thus very ``noise-sensitive''. (The noisy permanent
is simply the permanent multiplied by $\rho^n$.
In this example, while far apart, the permanent can be recovered perfectly from the noisy permanent.)
In this paper we study a closely related (but more interesting) example where
the function is the {\it square} of the permanent.

{\bf Remark:} Questions regarding noise sensitivity of various invariants of random matrices were raised by Itai Benjamini in the 
late 90s, see \cite{Kal00} Section 3.5.11. Kalai and Zeitouni proved \cite {KaZe07} that the event of having the largest 
eigenvalue of an $n$ by $n$ Gaussian matrix larger than (and also smaller than) its median value is noise sensitive.



\subsection {BosonSampling and Noisy Gaussian BosonSampling}

Quantum computers allow sampling from a larger class of probability distributions compared to classical randomized computers.
Denote by QSAMPLE the class of probability distributions that quantum computers can sample in polynomial time.
Aaronson and Arkhipov \cite {AaAr13}, and  Bremner, Jozsa, and Shepherd \cite {BJS11} proved that if QSAMPLE can be performed
by classical computers then
the computational-complexity polynomial hierarchy (PH, for short) collapses.
Aaronson and Arkhipov result applies already for BosonSampling. These important computational-complexity
results follow and sharpen older result by Terhal and DiVincenzo \cite {TeDi04}.

The main purpose of Aaronson and Arkhipov \cite {AaAr13} was
to extend these hardness results to account for the fact that implementations of quantum evolutions are noisy.
The novel aspect of \cite {AaAr13} approach was that they did not attempt to model the noisy evolution leading to the bosonic state but
rather made an assumption on the target state, namely that it is close in variation distance to the ideal state. They also considered the
case that the input matrix is Gaussian both because it is easier to create experimentally such bosonic states,
and because of computational complexity consideration.
They conjecture that approximate BosonSampling for random Gaussian input is already computationally
hard for classical computers (namely it already implies PH collapse), and show how
this conjecture can be derived from two other conjectures: A reasonable conjecture on the distribution of the permanents of
random Gaussian matrices together
with the conjecture that it is \#P hard to approximate the permanent
of a random Gaussian complex matrix.

Aaronson and Arkhipov proposed BosonSampling as a way to provide strong experimental
evidence that the ``extended Church-Turing hypothesis'' is false. Their hope is that current experimental methods
not involving quantum fault-tolerance may enable performing approximate BosonSampling for Gaussian
matrices for 10-30 bosons (``but not 1000 bosons'').
This range allows (exceedingly difficult) classical simulations and
thus the way quantum and classical computational efforts scale could be examined. ``If that can be done,'' argues Aaronson, ``it becomes harder
for QC skeptics to maintain that some problem of principle would inevitably prevent scaling to 50 or 100 photons.''

\subsection {Combinatorics of permutations and moments of permanents} 

A beautiful result by Aaronson and Arkhipov
asserts that for $n$ by $n$ complex Gaussian matrices\footnote{Aaronson and Arkhipov proved that the same formula 
holds for determinants and also studied higher moments.}

\begin{align}
\label {e:aa}
\Ex\left[|\perm (A)|^4\right] = (n+1)(n!)^2.
\end{align}

The proof of the complex case of our main theorem refines and
re-proves this result. It turns out that combinatorial argument
similar to the one used by Aaronson and Arkhipov is needed in the case
where $A$ is a real Gaussian matrix, to determine the contribution of
the top-degree Hermite coefficients of $|\perm (A)|^2$, and this
can then be used  to compute the contributions of all other degrees.






\section {Noise sensitivity - complex Gaussian matrices}
\label {s:complex}

In this section we analyse the permanent of an $n$ by $n$ complex
Gaussian matrix. We begin with a few  elementary definitions and
observations.

 \medskip 
We equip $\C^n$ with the
product measure where in each coordinate we have a 
Gaussian normal distribution with mean 0 and variance 1. We call a
random vector $z\in\C^n$ which is distributed according to this
measure a normal (complex) Gaussian vector. The measure also 
defines a natural  inner-product structure in the space of complex valued
functions on $\C^n$.

\paragraph{Noise operator and correlated pairs.} Let $\epsilon>0$ be a
noise parameter, let $\rho=\sqrt{1-\epsilon}$, and let $u$\gnote{why
  these letters?} be an independent Gaussian normal vector in
$\C^n$. For any $z\in\C^n$, we say that $y=\sqrt {1-\epsilon}\cdot z +\sqrt
\epsilon \cdot u$ is an $\epsilon$-noise of $z$. If $z$ is also a normal
Gaussian vector independent of $y$, we say that $y$ and $z$ are a
$\rho$-correlated pair.  For a function $f:\C^n\to\C$, we define the
noise operator $\T_\rho$ by \[T_\rho(f)(z)=\Ex[f(y)],\] where $y$ is
an $\epsilon$-noise of $z$.

\paragraph{An orthonormal set.} In order to study the noise
sensitivity of $\perm$, it is useful to use the following set of
orthonormal functions, related to the real Hermite basis. 
\begin {proposition} \label{prop:basic-complex}
The functions 1, $z$, $\bar z$ and $h_2(z)=z \bar z-1$ form an orthonormal set of functions. 
Moreover, these functions are all eigenvectors of $T_\rho$, with eigenvalues $1,\rho,\rho$ and $\rho^2$ respectively. 
\end {proposition} 
\newcommand{\barz}{{\bar z}}
\newcommand{\expect}{\mathbb E}
\paragraph{Proof.} The function $1$ obviously has norm $1$, and the
functions $z$ and $\bar z$ have norm $1$ since $z$ (and therefore
$\bar z$) have variance $1$. Also note that since $a=Re(z)$ and
$b=Im(z)$ are independent real normal variables  with expectation $0$
and variance $\frac12$, 
\begin{align*}
||z\barz||_2^2= \expect[|z|^4]=\expect[(a^2+b^2)^2]=\expect[a^4+b^2+2a^2b^2]=\frac34+\frac34+\frac12=2.
\end{align*}
Hence the norm of $h_2(z)$ is given by 
\begin{align*}
||z\cdot\bar z-1||_2^2=||z\bar z||_2^2 +1-2\langle z\barz, 1 \rangle = ||z\bar z||_2^2 +1-2\langle z,z \rangle =  2+1-2=1
\end{align*}

It is simple to verify that $1$, $z$, and $\barz$ are also all
orthogonal to each other (it follows since the Gaussian distribution
is symmetric around zero), and  that $z\barz-1$ is orthogonal to $1$. Also, 
$\langle z\barz-1,z \rangle=\expect[z\barz^2 -\barz]$, and the
expectations of both terms is again zero as they are odd functions of $z$.

It is left to show that the above functions are eigenvectors of $T_\rho$. This is obvious for $1$. 
For $f(z)=z$, $T_\rho(f)(z)=\expect[\rho z+\sqrt{1-\rho^2}u]=\rho z$, and similarly for $\barz$. Also,
\begin{align*}
T_\rho(h_2)(z)=&\expect[(\rho z+\sqrt{1-\rho^2}u )(\rho \barz +\sqrt{1-\rho^2}\bar u )]-1 
\\=&\rho^2z\barz+ \sqrt{1-\rho^2} \expect[z\bar u + \barz u] +\expect[(1-\rho^2)u\bar u] -1
 \\= &\rho^2z\barz+(1-\rho^2) -1=\rho^2\cdot h_2(z).
\end{align*}
\endproof

\newcommand{\comp}[1]{{#1}^c}
\paragraph{Permanents.} Let $\mathbf z=\{z_{i,j}\}_{i,j=1,\ldots,n}$
be an $n\times n$ matrix of independent complex Gaussians, 
and let $\perm(\mathbf z)= \sum_{\sigma\in S_n} \prod_{i=1}^n z_{i,\sigma(i)}$ be the permanent function. 
We also let 
\[f(\mathbf z)= |\perm (\mathbf z)|^2 = \sum_{\sigma,\tau\in S_n}
\prod_{i=1}^n z_{i,\sigma(i)}\barz_{i,\tau(i)}.\] In order to study
$T_\rho(f)$, consider one term in the formula above that corresponds
to the permutations $\sigma$ and $\tau$, and let $T$ be the indices
$i$ on which they agree, and $\comp T=[n]\setminus T$ be its complement. We can  write such a term as
\begin{align*}
\prod_{i=1}^n z_{i,\sigma(i)}\barz_{i,\tau(i)}&=\prod_{i\in
  T}(z_{i,\sigma(i)}\barz_{i,\sigma(i)}) \cdot \prod_{i\in \comp{T}}z_{i,\sigma(i)}\barz_{i,\tau(i)}
=\prod_{i\in T}(1+h_2(z_{i,\sigma(i)})) \prod_{i\in \comp{T}}z_{i,\sigma(i)}\barz_{i,\tau(i)} \\
&=\sum_{R\subseteq T}\left[\prod_{i\in T\setminus
    R}h_2(z_{i,\sigma(i)}) \prod_{i\in \comp T}z_{i,\sigma(i)}\barz_{i,\tau(i)} \right] 
\end{align*}
\paragraph{The degree of a term.}  For each product in the sum above
we assign a degree -- we add $1$ to the degree for each multiplicand
of the form $z_{i,j}$ or $\barz_{i,j}$, and $2$ for each multiplicand
of the form $h_2(z_{i,j})$. The degree of a term $\prod_{i\in
  T\setminus R}h_2(z_{i,\sigma(i)}) \prod_{i\in \bar
  T}z_{i,\sigma(i)}\barz_{i,\tau(i)}$ is thus
$2(|T|-|R|)+2(n-|T|)=2(n-|R|)$. 

\paragraph{The weight of $f$ on terms of degree $2(n-k)$.}  The
$2(n-k)$-degree part of $f$ is obtained by summing over all sets
$R\subseteq [n]$ of size $k$, the terms as above obtained from pairs
$(\sigma,\tau)$ of permutations which agree on the indices in $R$ (and
possibly on other indices).  It is useful to further partition these terms
according to the image $R'$ of $R$ under $\sigma$ and $\tau$ -- note
that there are $k!$ ways to fix the values of $\sigma$ and $\tau$ on
$R$ given $R'$. We denote by $\sigma',\tau'$ the restriction of
$\sigma$ and $\tau$ respectively on the complement of $R$, namely
these are one-to-one functions from $\comp R$ to $[n]\setminus
{R'}$. Also, let $S(\sigma',\tau')\subseteq \comp R$ be the set of indices
on which they agree.  So the degree $2(n-k)$ part of $f$ is given by
\begin{equation}\label{eq:poop}
  f^{=2(n-k)}=\sum_{|R|,|R'|=k} \left(
    k!\cdot\left[\sum_{\sigma',\tau'}\prod_{i\in S(\sigma',\tau')}
      h_2(z_{i,\sigma'(i)}) \prod_{i\in \comp R\setminus
        S(\sigma',\tau')} z_{i,\sigma'(i)}\barz_{i,\tau'(i)}
    \right]\right).
\end{equation}
Note that in the inner sum above no two summands are the same ($R$ and
$R'$, as well as $\sigma'$ and $\tau'$, can be inferred from  looking at
such a summand). Hence, since these summands form an orthonormal set, we have that
the weight of $f$ on its degree $2(n-k)$ terms is
\begin{equation}\label{eq:2}
||f^{=2(n-k)}||_2^2={n \choose k}^2\cdot(k!)^2\cdot((n-k)!))^2=(n!)^2,
\end{equation}
where the ${n \choose k}^2$ terms accounts for the possible values of
$R$ and $R'$, $(k!)^2$ comes from the coefficient of each summand
in~\eqref{eq:poop}, and $((n-k)!))^2$ is the number of choices for
$\sigma'$ and $\tau'$.

\medskip \noindent {\bf Remark:} Summing over all values of $k$, $1 \le k \le n+1$ we retrieve Aaronson and Arkhipov's formula (\ref {e:aa}).


\subsubsection*{Proof of Theorem \ref {thm:1} for the complex case}
Let $f,g,f'$ and $g'$ be as in Theorem~\ref{thm:1}, and recall that the correlation $corr(f,g)$ between $f$ and $g$ is given
by $corr(f,g)= <f',g'>/\|f'\|_2\|g'\|_2$. Also note that by the
definition of $T_\rho$, $g=T_\rho(f)$ for $\rho=\sqrt{1-\epsilon}$. 

\paragraph {The correlation diminishes when the noise is $\omega(1)/n$.}
It follows from
Proposition~\ref{prop:basic-complex} that the terms of degree $2m$ are
eigenvectors of the operator $T_\rho$ with eigenvalue
$\rho^{2m}$. 
We will use this observation together with \eqref{eq:2} to show that $corr(g,f)=o(1)$ when $\epsilon = \omega(1)/n$. 
Indeed, denoting $W_{2m}(n)=||f^{=2m}||_2^2$, we have 
$$\|f'\|_2=\parenth{ \sum_{m>0} W_{2m}(n) }^{1/2},$$
$$\|g'\|_2=\|T_\rho(f')\|_2=\parenth{ \sum_{m>0} W_{2m}(n)\rho^{4m} }^{1/2},$$
$$\inner{f',g'}=\sum_{m>0} W_{2m}(n)\rho^{2m}.$$
It follows that 
\begin{equation}
  \label{eq:3}
corr(f,g) = \frac{\sum_{m=1}^n \rho^{2m}}{
  (\sum_{m=1}^n 1)^{1/2} (\sum_{m=1}^n\rho^{4m})^{1/2}}\;.
\end{equation}
When $\epsilon=\omega(1)/n$, $\rho^2=1-\epsilon=1-\omega(1)/n$, and thus the enumerator in \eqref{eq:3} is of order
$\Theta(1/\epsilon)$ and the denominator is of order
$\Theta\parenth{\sqrt {n/\epsilon} }$. The correlation between $f$ and
$g$ in this case is therefore of order $\Theta\parenth{\sqrt {\epsilon
  n}}$, which indeed 
tends to zero when $\epsilon=\omega(1)/n$.

\paragraph{Proof of Corollary \ref {c:1}.} The corollary is obtained
from \eqref{eq:3} 
by using the formula for the
summation of a geometric series and the approximation $(1-\frac c
n)^n\sim \exp(-c)$.

\paragraph {Approximating the noisy permanent for a  constant noise
  parameter.} Note that the weight of the noisy permanent function,
$g$, on terms of degree $> d$, is bounded by $\rho^d\cdot
||g||_2^2$. Therefore $g$ can be approximated to within a $\rho^d\cdot
||g||_2^2$ distance by truncating terms of degree above $d$. 

It follows that when the noise parameter $\epsilon$ is constant, $g$
can be approximated to within any desired constant error by a linear
combination of terms each of degree  at most $d$. Moreover, as the
coefficient of each such term can be easily computed in polynomial
time, and since the number of such coefficient is a polynomial
function of $n$, this implies that $g$ can be approximated in
polynomial time up to any desired (constant) precision. 

This approximation of $g$ can even be achieved by a constant depth
circuit: this follows since each term, being of constant degree, can
be approximated to within polynomially small error in constant depth
as it only required taking $O(\log n)$ bits into account (it is
actually possible to only do computations over a constant number of
bits here by first applying some noise to the input variables). Then one can
approximate the sum of these terms by simply summing over a sample of
them, using binning to separately sample terms of different orders of
magnitude. We note that this argument is very general and only uses the fact
that $g$ can be approximated by an explicit constant degree
polynomial. 

\endproof

\subsection {Discussion}

\paragraph {Sharpness of the results.} Since our (Hermite-like) expansion of $|\perm^2(X)|$ is supported on degrees at most $2n$, we do 
have noise stability when the level of noise is $o(1/n)$. There is also a recent
result by Alex Arkhipov \cite {Ar14} that for certain general error-models, if the error per photon is  $o(1/n)$, 
``you'll sample from something that’s close in variation distance to the ideal distribution.'' (A careful comparison between 
Alex's result and ours shows that in our notions it applies when $\epsilon = o(1/n^2)$ leaving an 
interesting interval for noise-rate to be further explored.)
Independently from our work, 
Scott Aaronson \cite {Aa14} 
has a recent unpublished  (partially heuristic) result which shows that part (ii) 
of Theorem \ref {thm:1} is sharp for a different but related noise model:  
``Suppose you do a BosonSampling experiment with $n$ photons, suppose that $k$ out of the $n$ are 
randomly lost on their way through the beamsplitter network (you don't know which ones), 
and suppose that this is the only source of error.  Then you get a probability distribution that's hard to simulate 
to within accuracy $\theta (1/n^k)$ in variation distance, unless you can approximate the permanents of Gaussian matrices in $BPSUBEXP^{NP}$.'' 

\paragraph {Determinants.} We expect that our results apply to determinants and thus for FermionSampling 
and it would be interesting to work out the details. Perhaps a massage 
to be learned
is that the immense computational complexity gap between determinants and permanents is not manifested in the realistic 
behavior of fermions and bosons.\footnote{This is related to comments made by Naftali Tishby is the mid 90s. \cite{TrTi96}, however, proposes a physical distinction
between permanents and determinants in term of intrinsic variance of the measurement.}  Noise sensitivity gives an explanation why.



\paragraph {Permanents with repeated columns.}  
For the study of noise sensitivity of BosonSampling (when $m$ is not very large compared to $n$) we will need to 
extend our results to permanents of complex Gaussian matrices with repeated columns. This looks very interesting
and would hopefully be studied in a future work. Given an $n$ by $k$ matrix
$A=(z_{ij})_{1 \le i \le n, 1 \le j \le k}$, and $k$ integers $n_1,n_2\dots,n_k$ ,summing to $n$ we can let
$A'$ be the $n$ by $n$ matrix obtained by taking $n_i$ copies of column $i$
and define $f(A)= (1/n_1!n_2!\dots n_k! )\perm (A' A'^* )$. 
It is possible to expand $f(A)$ in a similar way to our computation above where only the 
combinatorics becomes somewhat more involved (and explicit formulas are not available). 
Of course, repeated columns are not relevant for FermionSampling.

\paragraph{BosonSampling: the normalization term.}

Given an $n$ by $m$ matrix $A=(z_{ij})_{1 \le i \le n, 1 \le j \le m}$ 
we will consider now the normalization term, $h$,  namely 
the $\mu(S)$-weighted sum of 
absolute value squared of permanents of all $n$ by $n$ minors. By the Cauchy-Binet formula for permanents \cite {Min78,HCB88},
\begin {align*}
h(A)= \perm(A A^*)=  \sum_{\sigma\in S_n}\sum_{k_1,k_2,..,k_n\in [m]}
\prod_{i=1}^n z_{i,k_i}\barz_{\sigma(i),k_i}.
\end {align*}

Again, it is possible to expand $h(A)$ in a similar way to our computation above. 
Of course, the (even more familiar) Cauchy-Binet theorem for determinants 
applies (in our setting) to the normalization term for FermionSampling.

\paragraph {Noise sensitivity for general polynomials in $z_i$ and $\bar z_i$}   
 
It will  be interesting to extend our framework and study noise sensitivity for 
general polynomials in $z_i$ and $\bar z_i$, or even just for absolute values of 
polynomials, parallel to \cite {BKS99} and \cite {KiOd12}. 
(This will be needed. e.g., for extensions of our results to higher moments of the complex Gaussian determinant and permanent.)

\paragraph{The Bernoulli case.} It will be interesting to prove similar results for other models of random matrices. 
A case of interest is when the entries of the matrix are i. i. d. Bernoulli random variables.   
To extend our results we need first to compute (or at least estimate)
the expectation of $|\perm(X)^4|.$
This is known for the determinant \cite {Tur55} (while more involved than the Gaussian case).




\section {Conclusion}
\label{s:conc}

Theorem \ref {thm:1} and its anticipated extensions propose the following 
picture: First, for constant noise level the noisy version of BosonSampling is in {\bf P}. In fact, noisy BosonSampling can be 
approximated by bounded depth circuits.
Second, when the level of noise is above $1/n$ 
when we attempt to approximate Gaussian bosonic states 
we cannot expect robust experimental outcomes at all. And third, when we consider perturbations of our Gaussian noise model, 
the noisy BosonSampling distribution will be very dependent on the 
detailed parameters describing the noise itself, so that for  robust outcomes, 
an exponential size input will be required to describe the noise.


The relevance of noise sensitivity may extend to more general quantum systems and 
this is an interesting topic for further research. 

\subsection*{Acknowledgment}
We would like to thanks Scott Aaronson, Alex Arkhipov, Micharl Ben-Or,
Michael Geller, Greg Kuperberg, Nadav Katz, 
Elchanan Mossell, and John Sidles, for helpful discussions.


\appendix

\section {Appendix 1: Modeling noise for BosonSampling}

A great advantage of Aaronson and Arkhipov's  BosonSampling proposal
is the simplicity, both of the ideal model, and
also of various noise models. 
In this section we will discuss some aspects of modeling noise for BosonSampling, starting with the rationale for the 
model we consider. 
Our model is motivated by a schematic picture for implementing BosonSampling based on 
creating separately $n$ photons in prescribed states, and reaching via interference 
a bosonic state for $n$ indistinguishable photons. For an individual photon we expect that 
our experimental process will lead to a mixture of (additive) Gaussian perturbations of the prescribed state.  
More importantly, we regard our simple model as relevant because 
we expect that the mathematical properties 
demonstrated here will extend to other modeling of noise.

\paragraph {How does a noisy single boson behave?}
One issue which is not addressed by us is that the amount of noise for achieving a single 
Boson with $m$ modes may also scale up with $m$.
The way noise scale up with the number of modes may depend on the state itself. 
We note that Krenn et als. \cite {KHF+13} were able to 
demonstrate a pair of entangled photons with $m=100$.


\paragraph {Other Noise models}

We are aware of a few other noise models that should be considered.

\begin {itemize} 

\item

Mode-mismatches. 
Mode mismatch means that photon detection is not
perfectly matched to photon states, so that the environment learns
something about the history of the observed photon.   As a result,
what was supposed to be two contributions to the quantum amplitude are
instead added as two probabilities. Mathematical modeling of mode-mismatches were offered by 
Charles Xu \cite {Xu13} and by Greg Kuperberg \cite {Ku14}. 
In Kuperberg's version if the ideal matrix is $M_{ij}$ then the noisy matrix 
is given as $M'_{ij} = exp(i\theta_{ij}) M_{ij}$, where $\theta_{ij}$ are i.i.d., and thus  mode
mismatch is described by i.i.d. noise in the phases of the matrix
entries. The modeling proposed by Xu and Kuperberg are mathematically 
similar with our model.

\item Multiplicative unitary noise.
When we think about the process of creating BosonSampling as unitary Gaussian operator
acting on $n$ bosons in an initial state, then it would be natural to consider mixture 
of the intended Gaussian operator 
with further {\em multiplicative} Gaussian-like unitary operator describing the noise. 

\item Inaccuracy of beamsplitters and phaseshifters. 
The photonic states are manipulated 
using beamsplitters and phaseshifters which pretty much have the roles of ``gates'' in 
the qubit/gate model of quantum computation. For a mathematical modeling
of noisy beamsplitters and phaseshifters and results of similar nature to ours 
see, e.g., Leverrier and Garcia-Patr\'on,\cite {LeGa13} 

\item  
Unheralded photon losses. This is a type of noise which is amply discussed 
in \cite {AaAr13} and subsequent works.

 
\item
Specific forms of noise for implementations of BosonSampling by superconducting or ion trapped qubits.
\end {itemize}

We expect that the noise sensitivity phenomenon and the suppression of high degree terms in a relevant 
Fourier-type expansion, 
will apply to {\em each one} of those forms of noise. (And also that quantitatively the effect of noise will be similar
to what we witness here.) 
The mathematics can be quite interesting and it will be interesting to explore 
it. Indeed our argument do apply (with small changes and an interesting combinatorial twist) 
to i. i. d. noise in the phases of the matrix entries.

\paragraph {Simulation}
It will be very interesting to make computer simulations to test how Gaussian noise of the kind we consider
here and other types of noise effect the permanent-squared and BosonSampling for small values of $n$ and $m$.
We expect that such simulations are pretty easy to implement and can be carried out for up to 15-20 bosons. 
It will also be interesting to compare the situation for permanents and determinants. 

When we consider specific implementation for BosonSampling we
may face the need for more detailed (and harder to implement) simulations.
We have learned from Nadav Katz and Michael Geller about some exciting implementation
of BosonSampling based on superconducting qubits and about detailed
simulations of these experiments. Those  simulations can be quite difficult
even for a few bosons, and simplified abstract modeling of noise of the kind proposed here
(and in Aaronson and Arkipov's papers, and the manuscripts by Kuperberg and Xu) 
can serve as intermediate steps towards a detailed and specific
modeling.

The difficulty in simulation of an experimental process
may give here and elsewhere an illusion of ``quantum supremacy,'' but we
have to remember that the primary obstacle for simulations is our
ability to understand and model the situation at hand, and that noise sensitivity suggests that modeling the situation at hand 
requires controlling exponentially many parameters.

\paragraph {Experimentation}
Of course, experiments will provide the ultimate test for BosonSampling. 
Indeed there are various remarkable experimental ways to go about it, 
either using ``photon machines,''or basing the implementation on highly stable qubits that are already possible 
via superconducting qubits or via ion traps. 
Here are a few references \cite {BFR+13,TDH+13,COR+13,SMH+13,KHF+13,SVB+14} \footnote {The first four cited simultaneous papers 
all appeared within two days on the archive!}. 
Our prediction regarding noise sensitivity could be tested in all these experimental implementation 
as well as with simulation based on information 
on the noise that can be based on experiments.

\section {Appendix 2:  Why BosonSampling may not work}
\label{appendix 2}


\subsection{How does realistic BosonSampling behave}

Our noise model is based on adding a random matrix with Gaussian entries. But there is no strong reasons to assume that 
the added random noise matrix will be so nicely behave. The space of $n$ by $m$ matrices is of dimension $nm$ and 
in the unit ball of probability distributions on this space we can find  a doubly exponential ``net'' of distributions 
such that each two have low correlation.  

Noise sensitivity for permanental-distributions proposes the following

1. Moving from one distribution of noisy matrices to another one which is 
$\Omega (1/n)$ apart (to be concrete, say, above $3/n$ apart in terms of correlation)
will lead with high probability to a small correlation (say, below 0.7 ) between the outcomes.

2. The size of a "net" of distributions which are  $3/n$-apart inside a ball of radius $
3/n$, is  doubly exponential in $n$. This continues to hold even if you impose further natural conditions on the 
distribution, such as statistical independence for the noise for different bosons.\footnote{If we allow undesirable interactions 
between the bosons this may increase
exponentially the dimension of the relevant Hilbert space 
 may lead to a net of triply-exponential size.})

This means that we may witness the following behavior:
\begin {itemize}

\item When the noise level is a constant 
then the resulting distribution will be classically simulable.
The asymptotic model describing the situation is polynomial 
and can be approximated by a (classical)  bounded-depth circuit. 

\item
 When the noise level $t$ is above $C/n$ 
getting a well defined distribution requires prescribing the noise,
which because of noise-sensitivity, depends on
an exponential input size.
From the point of view of Computational complexity, we  have an exponential
running time (with exponent $1/t$)
but exponential input size in $n$ as well. 
So no superior computational powers are manifested.

\item
In reality, even for a handful of bosons (7,8), 
it will simply not be possible to control or describe
 the noise in the required level to achieve a robust distribution.

\end {itemize}

\subsection {Noisy BosonSampling - computational complexity and practical reality}

While the specific relevance of noise sensitivity and the two barriers for noise-levels - $\omega(1)$ for computational 
feasibility and $\omega(1/n)$ for computational robustness  are novel, 
our point of view is overall consistent
with other researcher's viewpoint of BosonSampling. People do expect that, asymptotically, when $n$ is large, 
BosonSampling will require quantum fault-tolerance, and also the need for the noise to be below $1/n$ is consistent 
with earlier assertions (see, e.g. Leverrier and Garcia-Patr\'on  \cite {LeGa13}). The situation for BosonSampling is 
similar to what happens in standard, qubit-based quantum computing without fault-tolerance. 
Also there we can expect quantum fault-tolerance to be necessary even for implementing 
universal computation on a very small number of qubits.  
  
Still there is much hope among researchers that BosonSampling will be able to 
manifest ``quantum supremacy'' for 20 or even 30 Bosons. People do not see reasons why this cannot be achieved 
with current technologies. Moreover, there are several proposed avenues toward it. 
People see no obstacles for achieving it by traditional 
photonics and it can also be achieved via superconducting or ion trapped qubits. We note that those qubits can be created 
with fidelity levels approaching 99.99\% - which for many demonstrates that going below the $1/n$ barrier for dozens of bosons
is amply possible. Leverrier and Garcia-Patr\'on 
\cite {LeGa13} concluded that BosonSampling {\it is realistic} based on 
a similar barrier for the noise-level, since they did not think 
that this level is out of reach to experimentalists.
 
The missing part in the picture we draw is an explanation for why one can expect our picture to 
kick in for very few bosons (say, 8) rather than for a large number of bosons (say, 100). Of course, the best way to know is to experiment
and indeed we expect that for BosonSampling moving experimentally from three bosons to four and from four to five will be telling. 
Here we discuss  why the intuition that a ``constant level of noise''  or even polynomially small level 
of noise is ``just   
an engineering issue'' may be incorrect. 

We first point out a very simple but crucial computational theoretic insight: 
When we have a computational device (a noisy boson-sampler in our case) that when modeled formally 
cannot go (asymptotically) beyond {\bf P} 
then we usually 
should not expect it to be able to perform genuinely-hard computations (approximate BosonSampling in our case).

Noisy boson-Samplers as well as noisy Fermion-samplers  represent  a {\it very low} computational complexity class
(noisy polynomial-size bounded depth computation), which makes it less plausible that they will be algorithmically competitive in practice 
even to good classical algorithms.
   
This gives a clear computational-complexity based {\it reason} for why the task may well be out of reach to experimentalists. 

\subsection {The exponential curse for BosonSampling}

Let us go further to point out a sort of ``exponential explosion'' which 
characterizes the situation at hand. We already pointed out an ``exponential explosion'' for the number of 
parameters that may be needed to describe the noise, and we now mention a different related issue. 

The variety described by decomposable symmetric tensors inside the Hilbert space of
symmetric powers is of a very small dimension. It seems likely that as the parameters grow
our experimentally created bosonic states will not be confined or  close enough
to this variety.
We consider the variety of decomposable degree $n$ symmetric tensors with $m$ variables (of dimension $nm$ or so) 
inside the Hilbert space of all degree n symmetric tensors with m variables of dimension
$ {{n+m-1} \choose {n}}$. For example, , when $n=10,m=20$ we consider the 200-dimensional 
algebraic variety (of decomposable symmetric tensors parametrized by 10 by 20 complex matrices) inside a 20,000,000
dimensional Hilbert space (symmetric tensors). For 3 bosons the dimension of the variety is only roughly a third of that 
of the Hilbert space.\footnote{Of course, once we ``trace out'' the effect of the neglected parts
of the huge Hilbert space we may well end up with the type of noise considered here. 
So this item just gives a different point of view for the reason that the noise scales up and 
demonstrate the ``exponential curse'' that 
may obstruct BosonSampling already for few bosons.} In fact, since the relevant Hilbert space to start with is described 
by $n$ {\it distinguishable bosons}, its dimension $m^n$ is actually even much larger ($10^{13}$ for $n$=10, $m$=20).

\paragraph {The exponential curse and QC skepticism} 
The ``exponential curse,''namely,  the need to find a needle in an exponentially large haystack,  
is damaging for quantum computation as well as for classical computation. 
Error correction is a theoretical way around it. 
The first named author conjectures \cite {Kal11,KaHa12} 
that quantum error-correction and quantum fault-tolerance are not possible, 
and that the repetition mechanism (strongly related to the ``majority function\footnote{The main theorem of \cite {BKS99} gives an important 
connection between noise sensitivity and the majority function. It asserts that 
balanced Boolean functions which are not noise sensitive has substantial correlation with a weighted majority functions}
'') is the basis of any form of 
robust information and computation in nature. (Alas, only classical computation.) 

In other words, Kalai conjectures first 
that ``quantum supremacy'' 
requires quantum fault-tolerance, and second that quantum fault-tolerance is not possible. 
This paper supports the assertion that quantum supremacy requires quantum fault-tolerance.\footnote{It also supports the stronger 
conjecture that (quantum and classical) evolutions without fault-tolerance can be 
approximated by bounded-depth computation.}


\paragraph {Postselection}  The question if we can push down the noise level below the $1/n$ barrier for 20-30 bosons is mainly 
left to detailed experimentation, but if this {\it cannot be done}, noise-sensitivity gives gloomy prospects for methods 
based on postselection to tolerate larger rates of noise. For example, one postselection idea, referred to as Scattershot BosonSampling, 
is to have 200 imperfect sources for our photons, and then even if 
each source produce a photon with probability 10\%, we still be able to demonstrate BosonSampling distribution on the surviving 20 photons. 
Indeed you will not present the permanental distribution from a prescribed matrix but rather from an unknown-in-advanced submatrix, but this 
has no bearing on demonstrating ``quantum supremacy.'' Noise sensitivity suggests that no matter what the selected submatrix 
is the experimental outcomes are either meaningless or depend on an exponential number of parameters required to describe the noise.

\subsection {Varietal evolutions, varietal states and approximations}
Noise sensitivity and related insight on the spectral description of the effect of noise, 
can be relevant to the understanding of more general noisy quantum systems and we will indicate one direction.  
There is much implicit or explicit interest in quantum states which consist of low-dimensional algebraic variety and on
approximations to quantum evolutions (or quantum-like) evolutions
on such varieties.
It will be interesting to examine if our prediction that the noisy decomposable bosonic states
have good approximations in terms of ``low degree Hermite polynomials'' can be extended to general cases where
we reach states in low dimensional algebraic variety inside a high dimensional Hilbert space.  In other words, can
we identify the low dimensional Hilbert space directly in terms of the embedding of the variety.
Certainly, as we see from BosonSampling, the mere fact that we have a small-dimensional variety does 
not imply that polynomial-time approximations are possible.
It is possible that, in every such situation, small-degree polynomials in the the
tangent space to the variety allow already good approximation for {\it realistic} noisy quantum systems
which are approximately supported in such a variety. 
This will be a vast generalization
of our results and it will be interesting to explore it.

\subsection {The simulation heuristic for quantum speed-up proposals
which shortcut quantum fault tolerance}

BosonSampling is one of several proposals to shortcut quantum fault-tolerance in full or in part and still exhibit
quantum speed-up. The first-named author offered a general heuristic argument ``against'' such proposals:

\begin {itemize}
\item
You should be able to demonstrate the detailed/microscopic description of your experimental process
on a (hypothetical) noisy quantum computer without quantum fault-tolerance,
\end {itemize}

 or else

 \begin {itemize}
\item
 You should be able to manifest how quantum fault-tolerance is hidden in the experimental process.
\end {itemize}

This heuristic often suggests that experiments or
a detailed modeling on the proposed experimental
process (even with ordinary modeling of noise)
may be in conflict with the experimental hopes. (Of course, the heuristic
argument does not replace the need for such experiments or detailed modeling.)

The simulation heuristic can be applied for BosonSampling:
we can ask how errors scale up for a noisy quantum computer without fault-tolerance with noise tuned so that we
can create a single Gaussian boson state with $m$ modes with a fixed amount of noise,
when we move from one boson to to $n$-bosons states.
This poses a challenge for proponents of BosonSampling - to show how we can avoid scaling up the amount of noise with the number of bosons
when we simulate BosonSampling with noisy quantum circuits without the fault-tolerance apparatus.
The results in this paper give a more direct and stronger evidence compared to the simulation heuristic for this particular case.

\section{Appendix 3: Noise
sensitivity and robustness }

\subsection {Robust instances of noise-sensitive functions}

Noise sensitivity of BosonSampling leads to several questions in the theory of noise-sensitivity itself. 
We elaborate now on one such question.
There are robust bosonic states in nature and the discussion of noise-sensitivity of bosonic states
raises the following general question for noise-sensitivity.


\begin {quotation}
{\bf Problem:}
Understand noise stable instances of noise-sensitive functions.
\end {quotation}

A related interesting question is:

\begin {quotation}
{\bf Problem:}
Understand noise sensitive instances of noise-stable functions.
\end {quotation}



\paragraph{ Percolation} Consider the crossing event in planar percolation on $n$ by $n$ square grid. Benjamini, Kalai and Schramm \cite {BKS99} proved that this function is noise sensitive
and very strong form of noise sensitivity were subsequently proved by Schramm and Steif \cite {ScSt10}, and Garban, Pete and Schramm \cite {GPS10}, see also \cite {GaSt14}. It is an interesting question
to identify cases where the crossing event is robust. Of course, a choosing an edge to be open with probability $p>1/2$ (independently) will give you with high probability such a robust crossing event. Another example is
to consider $X$ - the $\log n$ neighborhood of a left-right crossing, and take every edge in $X$ with probability $p>1/2$ (independently). It will be interesting to describe all stable-under-noise
crossing states.

{\bf Tribes and recursive majority.} Those are well known simpler noise-sensitive functions \cite {BL90,KKL88,BKS99} where the situation may be easier. 
Robust states for the tribe function can perhaps be described easily. 
We can define
for a $\pm 1$-vector the fraction $u(t)$ of tribes where more than a fraction of $t$ of the variables are equal to one. It looks that for a 
level of noise $\rho$ (asymptotically as $n$ grows)the robustness of a state is determined by this function. But maybe there are robust states of other kind. It will be 
interesting to identify the
robust
instances for the recursive ternary majority which is another basic example of noise sensitive Boolean function.

\paragraph {Squares of permanents and bosonic states.} It will be of much interest to identify

\begin {quotation}
{\bf Problem:} Describe $n$ by $n$ complex matrices, and bosonic states that are noise sensitive,
namely so that the noisy value/distribution (obtained by taking the expectation after adding a Gaussian noise)
is close to the original value/distribution.
\end {quotation}



{\bf Remark:} It is an interesting question which bosonic states are realistic and noise
stability can be relevant to the answer.
Flammia and Harrow \cite {FlHa13} used certain bosonic states to disprove
a proposed criterion of Kalai \cite {Kal11} for ``non physical'' quantum states.

\paragraph {FourierSampling and anyons} FourierSampling is among the most useful quantum subroutines. 
We can ask about noise sensitivity of FourierSampling, and about robust states for FourierSampling.

Anyons of various types are also important for quantum computing and we can ask about noise-sensitivity of various anyonic
states. An important difference between anyons and bosons/fermions is that we do not have the analog
of ``decomposable'' states (those which as symmetric tensors have rank-1 and are thus described based on minors of a single matrix).







\section {Appendix 4: The power of quantum sampling compared to BQP.} 

One of the fascinating aspects of the
study of probability distributions that can be achieved efficiently by quantum computers
is that it is possible that the computation power of quantum computers for sampling is much stronger than the computational
advantage they have for decision problems.


\begin {quotation}

{\bf Problem:} (\cite{Kal10})
Does the assumption that a classical computer with BQP subroutine can perform
QSAMPLING (or just BosonSampling or FourierSampling)
already leads to polynomial-hierarchy collapse or other
computational complexity consequences of a similar nature?

\end {quotation}

\section {Appendix 5: noise sensitivity and permanents - real Gaussian matrices}
\label {s:real}

\paragraph {proof of Theorem \ref {thm:1} (real case)}

\paragraph{Hermite polynomials} We do our computations in terms of
Hermite polynomials. Here are the facts that we use:
 The univatiate Hermite polynomials $\set{h_d}_{d=1}^\infty$
are have norm $1$, they are orthogonal
w.r.t. the Gaussian measure, and also $h_d$ is of degree $d$. This
defines them uniquely. The degree $0$ and degree $1$ normalized Hermite
polynomials in $x$ are $h_0(x)=1$ and $h_1(x)=x$ respectively: it is
easy to verify that they have norm $1$ and that they are
orthogonal. It is also easy to see that
$h_2(x)=\frac{1}{\sqrt2}\cdot(x^2-1)$ is the normalized degree-$2$
Hermite polynomial: it is of the right degree and clearly orthogonal
to the first two polynomials. To verify that the norm is $1$ one only
needs to know that $\Ex[x^4]=3$ for a normalized Gaussian variable
$x$.



\paragraph{The Hermite expansion of the permanent squared} Recall that the
permanent of $X$ is a sum of products over all permutations in $X$,
and thus the square of the permanent is given by
\[
f=\perm(X)^2=\sum_{\tau,\sigma}\prod_{i=1}^n X_{i,\tau(i)}\cdot X_{i,\sigma(i)},
\]
where $\tau$ and $\sigma$ are permutations. 
To compute the expansion in terms of Hermite polynomials we consider first 
the contribution of a single pair
$(\tau,\sigma)$ of permutations. Let $T=\{i\in [n]: \sigma(i)=tau(i)\}$. $T=|FP(\sigma^{-1}\tau)|$ 
where $FP(\pi)$ is the set of fixed points of $\pi$.

\begin{align}
\label{eq:1}
\prod_i X_{i,\tau(i)}\cdot X_{i,\sigma(i)} &= \prod_{i\in T}
\parenth{(1+\sqrt 2)\cdot h_2(X_{i,\tau(i)})}\cdot \prod_{i\in
  [n]\setminus T}\parenth{X_{i,\tau(i)} X_{i,\sigma(i)}}  
\end{align}

$$= \sum_{S\subseteq T} \Brac{ \cdot 2^{|S|/2} \cdot \prod_{i\in S}
 h_2(X_{i,\tau(i)})\cdot \prod_{i\in
  [n]\setminus T}X_{i,\tau(i)} X_{i,\sigma(i)}    }\ .$$

Note that in equation (\ref{eq:1}) the same Hermite polynomial can come from 
different pairs of permutations. Let $W_k$ be the sum of squares of degree $k$ coefficients
in the Hermite expansion of $f$. We denote by $W_{2k}(n)$ the sum of squares of Hermite coefficients for Hermite monomials of degree $2k$.

\paragraph{The degree $2n$ contributions} 
We use the combinatorial identity $\sum_{\pi \in S_n}2^{cyc
  (\pi)}=(n+1)$.  The top degree $2n$ contribution accounts for the
case that $S=T$.  For a permutation $\pi \in S_n$ let $cyc(\pi)$
denote the number of cycles of $\pi$ (in its representation as the
product of disjoint cycles), and $cyc_{\ge 2}(\pi)$ denote the number
of cycles of size at least 2.  Note that the Hermite monomials of
degree $2n$ correspond to the set $M$ of pairs
$\{(i,\sigma(i)),(i,\tau(i)):i=1,2,\dots,n\}$. Let $\cal M$ denote the
set of all such $M$s.  The number of pairs of permutations that
correspond to the same $M$ is $2^{cyc_{\ge 2}(\sigma^{-1} \cdot
  \tau)}.$ Thus we have

\begin{align}
\label{e:expanW}
W_{2n}(n)=\sum_{M \in {\cal M}}2^{|FP(\sigma^{-1}\tau)|}4^ {cyc_{\ge 2}(\sigma^{-1} \cdot \tau)}= 
\end {align}

$$=\sum_{\sigma,\tau \in S_n}2^{|FP(\sigma^{-1}\tau)|}4^ {cyc_{\ge 2}(\sigma^{-1} \cdot \tau)}2^ {-cyc_{\ge 2}(\sigma^{-1} \cdot \tau)}=$$

$$=\sum {\sigma,\tau \in S_n}2^{|FP(\sigma^{-1}\tau)|}2^ {cyc(\sigma^{-1} \cdot \tau)}=(n!)^2(n+1).$$

\paragraph{The degree $2m$ contributions} 

Let $m-n-s$, degree $2m$ coefficients represent the terms in equation (\ref {eq:1}) 
contributed by sets $S$ with $|S|=s$. 
We have $n \choose s$ ways to choose $S$ 
and $n \choose s$ ways to choose $\tau(S)$. Given $S$ and $\tau(S)$, the same argument we used for 
equation (\ref{e:expanW}), gives that the sum of squares of the Fourier coefficients is $(s!)^2W_{2m}(m)$. 
(The term $(s!)^2$ accounts for all bijections from $S$ to $\tau (S)$ which all contributes to the same Hermite term.) This gives

\begin{align}
\label{e:expanW2}
W_{2m}(n)= {n \choose s}^2 (s!)^2 (n-s!)^2(m+1) = (n!)^2 (m+1).
\end {align}

Adding up the contributions of the different degrees  we get that for real Gaussian matrices 
$\|f\|_2^2=\mathbb E |(\perm (A)|^4 = {{n+2}\choose {2}}(n!)^2.$ 
(This also follows directly from the argument in \cite {AaAr13}, taking into account that the 4th moment of a standard 
real normal variable is 3 and not 2 as in the complex case.) 
The conclusions of both parts of Theorem \ref {thm:1} remain valid ) 
Now, both parts (i) and (ii) of Theorem \ref {thm:1} follows easily from relation (\ref{e:expanW2}).

\paragraph {The correlation diminishes when the noise is $\omega(1)/n$.}
The correlation $corr(f,g)$ between $f$ and $g$ is given by $corr(f,g)= <f',g'>/\|f'\|_2\|g'\|_2$.
We will use equation (\ref{e:nf}) to show that $corr(g,f)=o(1)$ when $\rho = \omega(1)/n$. Indeed,
$$\|f\|_2=(\sum_{m>0} W_{2m}(n))^{1/2},$$
$$\|g\|_2=\|T_\rho(f)\|_2=(\sum_{m>0} W_{2m}(n)(1-\rho)^m)^{1/2},$$
$$<f,g>=<f,T_\rho(f)>=(\sum_{m>0} W_{2m}(n)(1-\rho)^m.$$
It follows that $$corr(f,g) = \sum_{m=1}^n (m+1)(1-\rho)^m/ (\sum_{m=0}^n(m+1))^{1/2} (\sum_{m=1}^n(m+1)(1-\rho)^m)^{1/2},$$ which indeed 
tends to zero when $\rho=\omega(1)/n$.

\paragraph {The noisy state in in P when the noise is a constant.}
When the noise level is slightly above $1/d$, $g$ is well approximated by the truncation of the Hermite expansion 
for degrees at most $d$. We have polynomially many coefficient and it 
is easy to see that each coefficient requires a polynomial time computation.   


$\square$.

\end{document}